\documentclass{ws-ijgmmp}

\begin{document}

\markboth{Binaya K. Bishi, S.K.J. Pacif, P.K. Sahoo, G. P. Singh}
{LRS Bianchi type-I cosmological model with constant deceleration parameter in $f(R,T)$ gravity}

\catchline{}{}{}{}{}
%
%%%%%%%%%%%%%%%%%%%%%%%%%%%%%%%%%%%%%%%%%%%%%%%%%%%%%%%%%%%%%%%%%%%%

\title{\textbf{LRS Bianchi type-I cosmological model with constant deceleration parameter in $f(R,T)$ gravity}}

\author{Binaya K. Bishi}

\address{Department of Mathematics,\\ Visvesvaraya National Institute of Technology,\\ Nagpur-440010,
India.\\
\email{binaybc@gmail.com}}

\author{S.K.J. Pacif}

\address{Centre for Theoretical Physics,\\ Jamia Millia Islamia,\\ New Delhi-110025,
India.\\
\email{shibesh.math@gmail.com}}

\author{P.K. Sahoo}

\address{Department of Mathematics,\\ Birla Institute of Technology and Science-Pilani,\\
Hyderabad Campus,\\
Hyderabad-500078, India\,\\
\email{pksahoo@hyderabad.bits-pilani.ac.in} }

\author{G. P. Singh}

\address{Department of Mathematics,\\ Visvesvaraya National Institute of Technology,\\ Nagpur-440010,
India.\\
\email{gpsingh@mth.vnit.ac.in}}

\maketitle

\begin{history}
\received{(09 May 2017)}
\revised{(21 June 2017)}
\end{history}

\begin{abstract}
A spatially homogeneous anisotropic LRS Bianchi type-I cosmological model is
studied in $f(R,T)$ gravity with a special form of Hubble's parameter, which
leads to constant deceleration parameter. The parameters involved in the
considered form of Hubble parameter can be tuned to match, our models with
the $\Lambda $CDM model. With the present observed value of the deceleration
parameter, we have discussed physical and kinematical properties of a
specific model. Moreover, we have discussed the cosmological distances for
our model.
\end{abstract}



\section{Introduction}
Observation plays a major role in modern cosmology. The advent of new
technologies in observations enforces the theorists to rethink on the
formulation of the gravitational theories time to time. Einstein had to drop
the cosmological constant from the field equations with the discovery of
Hubble. The concept of decelerating expansion of the Universe had to drop by
the theorists with the observation of type Ia supernovae in 1998. Since then
CMB, BAO, SDSS and many more observations provide evidences in support of
the accelerating expansion of the Universe. So, it is very important to take
care of the observational results while building a theoretical model of the
Universe. The accelerating expansion of the Universe is an important feature
of present day cosmology. The Einstein field equations (EFEs) always lead to
a decelerating expansion with the normal matter component in the Universe.
The accelerating expansion can be described either by supplying some extra
component in the energy momentum tensor part in the field equations or by
doing some modifications in the geometrical part. With these principles, the
past few years of research produced a plethora of cosmological models of the
Universe explaining the accelerating expansion. The theory of \textit{dark
energy} have taken special status in recent times. The dark energy is an
exotic energy component with negative pressure, which explain many
observations well and solves some major problems of standard cosmology. The
second possibility is by assuming that the general relativity breaks down at
large scales and the gravitational field can be described by a more general
action.

The $f(R)$ theory of gravity \cite{Capozziello2005, Nojiri2006, Nojiri/2006, Nojiri2008, Nojiri/2011, Capozziello/2011, Nojiri/2017}
is an alternative to General Relativity (GR) to justify the cosmic
acceleration and early inflation in different way. In $f(R)$ theory, the
cosmic acceleration is obtained by the term $\frac{1}{R}$ where $R$ is the
Ricci scalar in the Einstein-Hilbert action. $f(R)$ gravity models also
addressed the issue of dark matter \cite{Hu2007, Appleby2007,
Starobinsky2007, Nojiri2007}. Recently, the mimetic $F(R)$ gravity \cite{Nojiri/2014, Odintsov/2015, Odintsov/2016} has been proposed to investigate the early-time and late-time acceleration of the universe. It is demonstrated that the mimetic $F(R)$ gravity consistent with Plank and BICEP2/Keck Array observations. The $f(R)$ gravity was modified by introducing
the trace of energy momentum tensor $T$ to the action yielding $f(R,T)$
gravity \cite{Harko2011}. The action for the $f(R,T)$ gravity is given as
\begin{equation}
S=\int \sqrt{-g}\left( \frac{f(R,T)}{16\pi}+L_{m}\right) d^{4}x  \label{1}
\end{equation}%
where $L_{m}$ is the matter Lagrangian and $g=|g_{ij}|$. Varying the action
in equation (1) with respect to metric tensor $g_{ij}$ the field equations
are obtained as
\begin{multline}
f_{R}(R,T)R_{ij}-\frac{1}{2}f(R,T)g_{ij}+(g_{ij}\Box -\nabla _{i}\nabla
_{j})f_{R}(R,T)=\\ 8\pi T_{ij}-f_{T}(R,T)T_{ij}-f_{T}(R,T)\Theta _{ij}
\label{2}
\end{multline}%
where
\begin{equation}
\Theta _{ij}=-2T_{ij}+g_{ij}L_{m}-2g^{lm}\frac{\partial ^{2}L_{m}}{\partial
g^{ij}\partial g^{lm}}  \label{3}
\end{equation}%
Here $f_{R}(R,T)=\frac{\partial f(R,T)}{\partial R}$, $f_{T}(R,T)=\frac{%
\partial f(R,T)}{\partial T}$, $\Box \equiv \nabla ^{i}\nabla _{i}$ where $%
\nabla _{i}$ represents covariant derivative.\newline
Contraction of equation (\ref{3}) yields
\begin{equation}
f_{R}(R,T)R+3\Box f_{R}(R,T)-2f(R,T)=(8\pi -f_{T}(R,T))T-f_{T}(R,T)\Theta
\label{4}
\end{equation}%
where $\Theta =g^{ij}\Theta _{ij}$. From equations (\ref{2}) and (\ref{4}),
one can obtain
\begin{multline}
f_{R}(R,T)\biggl(R_{ij}-\frac{1}{3}Rg_{ij}\biggr)+\frac{1}{6}f(R,T)g_{ij}=\\
8\pi -f_{T}(R,T)\biggl(T_{ij}-\frac{1}{3}Tg_{ij}\biggr) 
-f_{T}(R,T)\biggl(\Theta _{ij}-\frac{1}{3}\Theta g_{ij}\biggr)+\nabla
_{i}\nabla _{j}f_{R}(R,T)\text{.}  \label{5}
\end{multline}%
\qquad

Numerous works have been done in the past few years in $f(R,T)$ theory of
gravity due to the growing interests on the modified theories. One can see a
recent work for a case study on $f(R,T)$ gravity in Salehi and Aftabi \cite%
{Salehi2016}. Hundjo et al. \cite{Houndjo12} has developed the cosmological
reconstruction of $f(R,T)$ gravity and discussed the transition of matter
dominated phase to an accelerated phase. The non-equilibrium picture of
thermodynamics at apparent horizon for Friedmann-Robertson-Walker (FRW) universe is discussed in this
theory \cite{Sharif2012}. Sharif et al. \cite{Sharif2013} have studied
various energy conditions in $f(R,T)$ gravity and they reduce the same to $%
f(R)$ and $f(T)$ gravity. The Godel solutions are derived in this modified
theory \cite{Santos2013, Santos2015}. Memoni et al. \cite{momenietal16} have
studied the generalized second law of thermodynamics in $f(R,T)$ gravity.
The effect of bulk viscosity in $f(R,T)$ gravity is discussed for FRW metric
\cite{kumarandsingh15}. Shri Ram and Chandel \cite{RamandChandel15} have
discussed dynamics of magnetized string cosmological model. Two classes of $%
f(R,T)$ gravity models is investigated by Shamir and Raja \cite%
{shamirandraja15} for cylindrically symmetric space-time. Mores et al. \cite%
{moracesetal16} have discussed about the hydro static equilibrium
configuration of neutron stars and strange stars in the contexts of $f(R,T)$
gravity. Here the fluid pressure is computed from the equations of state (EoS) $%
\rho =\omega \rho ^{\frac{5}{3}}$ and $p=0.28(\rho -4B)$, where $B$ is a
constant and $\rho $ is the energy density of the fluid. Alhamzawi and
Alhamzawi \cite{AlhamzawiandAlhamzawi16} have discussed the gravitational
lensing in first class of $f(R,T)$ gravity. They have calculated the effect
of $f(R,T)$ gravity on gravitational lensing and shown that it can give a
considerable contribution to gravitational lensing. Mores \cite%
{moracesetal16a} has discussed the varying speed of light in $f(R,T)$
gravity. Alves et al. \cite{Alves2016} have studied the gravitational waves
scenario in this theory. Yousaf et al \cite{Yousaf2016} have investigated
the irregularity factor of self gravitating star due to imperfect fluid in $%
f(R,T)$ gravity.

Though the observations is in favour of a homogeneous and isotropic
Universe, the possibility of anisotropic phase in the early Universe is also
supported by some observations. Also the presence of anisotropy affect the
evolution of energy density. Evolution of anisotropic source for axially
symmetric universe have been discussed in $f(R,T)$ gravity \cite{Zubair2015}. The dynamical analysis of anisotropic spherically symmetric collapsing
star has presented in this modified gravity \cite{Noureen2015}. This
motivates the theorists to construct various models in different Bianchi
space-times in different contexts \cite{Readyandkumar13, Sahoo2015,
singhandbishi15, Sofuoglu16, Sahoo2017, Sahoo/2017}. In this paper, we consider LRS
Bianchi-I space-time as our background metric and study the evolution of
various cosmological parameters in $f(R,T)$ theory of gravity.

\section{Metric and Field Equations for $f(R,T)=f_1(R)+f_2(T)$}

The spatially homogeneous anisotropic LRS Bianchi type-I metric
\begin{equation}
ds^{2}=dt^{2}-A(t)^{2}(dx^{2}+dy^{2})-B(t)^{2}dz^{2}\text{,}  \label{6}
\end{equation}%
is symmetric corresponding to $xy-$plane. The average scale factor $a$,
spatial volume $V$, scalar expansion $\theta $ for metric (6) are
\begin{equation}
a=(A^{2}B)^{\frac{1}{3}},\ \ V=a^{3}=A^{2}B,\ \ \theta =u_{;i}^{i}=2\frac{%
A^{\prime }}{A}+\frac{B^{\prime }}{B}\text{.}  \label{7}
\end{equation}%
For $f(R,T)=f_{1}(R)+f_{2}(T)$, we consider the linear form $%
f_{1}(R)=\lambda R$ and $f_{2}(T)=\lambda T$ where $\lambda $ is an
arbitrary constant. Hence, in this case $f(R,T)=\lambda (R+T)$. We
considered the source of matter as perfect fluid having energy momentum
tensor
\begin{equation}
T_{ij}=(\rho +p)u_{i}u_{j}-pg_{ij}  \label{8}
\end{equation}%
where $u^{i}=(0,0,0,1)$ is four velocity vector satisfying $u^{i}u_{i}=1$,
the $f(R,T)$ gravity field equations (\ref{5}) takes the form
\begin{equation}
\lambda R_{ij}-\frac{1}{2}\lambda (R+T)g_{ij}+(g_{ij}\Box -\nabla _{i}\nabla
_{j})\lambda =8\pi T_{ij}-\lambda T_{ij}+\lambda (2T_{ij}+pg_{ij})\text{.}
\label{9}
\end{equation}%
Since $(g_{ij}\Box -\nabla _{i}\nabla _{j})\lambda =0$, we obtain
\begin{equation}
R_{ij}-\frac{1}{2}Rg_{ij}=\biggl(\frac{8\pi +\lambda }{\lambda }\biggr)%
T_{ij}+\biggl(p+\frac{1}{2}T\biggr)g_{ij}\text{ .}  \label{10}
\end{equation}%
From GR the Einstein tensor $G_{ij}\equiv R_{ij}-\frac{1}{2}g_{ij}R$. Using
this in equation (\ref{10}), we can write
\begin{equation}
G_{ij}-\biggl(p+\frac{1}{2}T\biggr)g_{ij}=\biggl(\frac{8\pi +\lambda }{%
\lambda }\biggr)T_{ij}.  \label{11}
\end{equation}%
In GR, the field equations with cosmological constant $\Lambda $ usually
written as
\begin{equation}
G_{ij}-\Lambda g_{ij}=-8\pi T_{ij}\text{ .}  \label{12}
\end{equation}%
Here, we assume a small $-$ve value for $\lambda $ throughout the manuscript
to get a better analogy with usual Einstein field equations.\newline
Comparison of (\ref{11}) and (\ref{12}) gives us
\begin{equation}
\Lambda \equiv \Lambda (T)=p+\frac{1}{2}T\text{,}  \label{13}
\end{equation}%
and $\lambda =-\frac{8\pi }{8\pi +1}$. In other words, $p+\frac{1}{2}T$
behaves as cosmological constant. The field equations (\ref{10}), for the
metric (\ref{6}) can be obtained as
\begin{equation}
\frac{A^{\prime \prime }}{A}+\frac{A^{\prime }B^{\prime }}{AB}+\frac{%
B^{\prime \prime }}{B}=\biggl(\frac{8\pi +\lambda }{\lambda }\biggr)%
p-\Lambda \text{,}  \label{14}
\end{equation}%
\begin{equation}
2\frac{A^{\prime \prime }}{A}+\biggl(\frac{A^{\prime }}{A}\biggr)^{2}=\biggl(%
\frac{8\pi +\lambda }{\lambda }\biggr)p-\Lambda \text{,}  \label{15}
\end{equation}%
\begin{equation}
\biggl(\frac{A^{\prime }}{A}\biggr)^{2}+2\frac{A^{\prime }B^{\prime }}{AB}=-%
\biggl(\frac{8\pi +\lambda }{\lambda }\biggr)\rho -\Lambda \text{,}
\label{16}
\end{equation}%
where an overhead prime denote derivative with respect to time `$t$' only.
The trace $T$ for this model is $T=-3p+\rho $, so that equation (\ref{13})
reduces to
\begin{equation}
\Lambda (T)=\frac{1}{2}(\rho -p)\text{.}  \label{17}
\end{equation}%
From equations (\ref{14}) and (\ref{15}), we have
\begin{equation}
\frac{A^{\prime }}{A}-\frac{B^{\prime }}{B}=\frac{c_{1}}{A^{2}B}\text{,}
\label{18}
\end{equation}%
where $c_{1}$ is constant of integration. Again integrating
\begin{equation}
\frac{A}{B}=c_{2}\exp \biggl[c_{1}\int \frac{dt}{A^{2}B}\biggr]=c_{2}\exp %
\biggl[c_{1}\int \frac{dt}{a^{3}}\biggr]\text{,}  \label{19}
\end{equation}%
where $c_{2}$ is integration constant.\newline
Using the above value in equation (\ref{7}), we can get
\begin{equation}
A=c_{2}^{1/3}a\exp \biggl[\frac{c_{1}}{3}\int \frac{dt}{a^{3}}\biggr]
\label{20}
\end{equation}%
and
\begin{equation}
B=c_{2}^{-2/3}a\exp \biggl[\frac{-2c_{1}}{3}\int \frac{dt}{a^{3}}\biggr]%
\text{.}  \label{21}
\end{equation}%
The directional Hubble parameters are defined as $H_{1}=\frac{A^{\prime }}{A}
$ and $H_{2}=\frac{B^{\prime }}{B}$ comes out as $H=\frac{1}{3}%
(2H_{1}+H_{2}) $ and $\theta =3H$. The shear scalar $\sigma ^{2}$ for the
metric (\ref{6}) is written as
\begin{equation}
\sigma ^{2}=\frac{1}{2}\biggl[\sum H_{i}^{2}-\frac{1}{3}\theta ^{2}\biggr]=%
\frac{1}{3}(H_{1}-H_{2})^{2}\text{.}  \label{22}
\end{equation}%
Using directional Hubble parameters, we can write the field equations (\ref%
{14})-(\ref{16}) as
\begin{equation}
H_{1}^{\prime }+H_{2}^{\prime }+H_{1}^{2}+H_{2}^{2}+H_{1}H_{2}=\alpha
p-\Lambda \text{,}  \label{23}
\end{equation}%
\begin{equation}
2H_{1}^{\prime }+3H_{1}^{2}=\alpha p-\Lambda \text{,}  \label{24}
\end{equation}%
\begin{equation}
H_{1}^{2}+H_{1}H_{2}=-\alpha \rho -\Lambda \text{,}  \label{25}
\end{equation}%
where $\alpha =\frac{8\pi +\lambda }{\lambda }$. The Ricci scalar $R$ for
our model is
\begin{equation}
R=-2\biggl[2\frac{A^{\prime \prime }}{A}+2\frac{A^{\prime }B^{\prime }}{AB}+%
\frac{B^{\prime \prime }}{B}+\biggl(\frac{A^{\prime }}{A}\biggr)^{2}\biggr]
\label{26}
\end{equation}%
Pressure, energy density and the cosmological constant for the model can be
written in terms of Hubble parameter as
\begin{equation}
p=\frac{(4\alpha +2)H_{1}^{\prime }+(6\alpha +2)H_{1}^{2}-H_{1}H_{2}}{%
2\alpha (\alpha +1)}  \label{27}
\end{equation}%
\begin{equation}
\rho =\frac{2H_{1}^{\prime }+(2-2\alpha )H_{1}^{2}-(2\alpha +1)H_{1}H_{2}}{%
2\alpha (\alpha +1)}  \label{28}
\end{equation}%
\begin{equation}
\Lambda =-\frac{2H_{1}^{\prime }+4H_{1}^{2}+H_{1}H_{2}}{2(\alpha +1)}
\label{29}
\end{equation}%
The equation of state parameter i.e. the ratio between pressure and energy
density is
\begin{equation}
\omega =\frac{(4\alpha +2)H_{1}^{\prime }+(6\alpha +2)H_{1}^{2}-H_{1}H_{2}}{%
2H_{1}^{\prime }+(2-2\alpha )H_{1}^{2}-(2\alpha +1)H_{1}H_{2}}  \label{30}
\end{equation}

Having a general set up, we look for solutions to the field equations in the
next section.

\section{Solution of the Field Equations}

In order to obtain an explicit solutions to the field equations, we require
a supplementary constrain equation for the consistency of the system. This
one extra constrain can be chosen by assuming linear relationship between
two variables in the field equations or we can parametrize any particular
variable. For a recent review on various parametrization one can see \cite%
{Pacif2017}. Recently Pacif and Mishra \cite{PacifMishra15} have proposed
special law of variation of Hubble parameter
\begin{equation}
H=\frac{m}{k_{1}t+k_{2}},  \label{31}
\end{equation}%
where $m\geq 0$, $k_{1}\neq 0$ and $k_{2}$ are constants and this readily
gives the scale factor explicitly as
\begin{equation}
a(t)=k_{3}(k_{1}t+k_{2})^{\frac{m}{k_{1}}}\text{,}  \label{32}
\end{equation}%
where $k_{3}$ is integration constant. The deceleration parameter $q$ comes
out to be a constant depending on $k_{1}$and $m$.
\begin{equation}
q=-1+\frac{d}{dt}\left( \frac{1}{H}\right) =-1+\frac{k_{1}}{m}\text{.}
\label{33}
\end{equation}%
Using the equation (\ref{32}) in (\ref{20}) and (\ref{21}) the metric
potentials are obtained as functions of time as
\begin{equation}
A=c_{2}^{\frac{1}{3}}k_{3}(k_{1}t+k_{2})^{\frac{m}{k_{1}}}\times \exp \left[
\frac{c_{1}(k_{1}t+k_{2})^{1-\frac{3m}{k_{1}}}}{3k_{3}^{3}(k_{1}-3m)}\right]
\label{34}
\end{equation}%
\begin{equation}
B=c_{2}^{\frac{-2}{3}}k_{3}(k_{1}t+k_{2})^{\frac{m}{k_{1}}}\times \exp \left[
\frac{-2c_{1}(k_{1}t+k_{2})^{1-\frac{3m}{k_{1}}}}{3k_{3}^{3}(k_{1}-3m)}%
\right]  \label{35}
\end{equation}%
The directional Hubble parameters $H_{1}$ and $H_{2}$ becomes
\begin{equation}
H_{1}=\frac{c_{1}(k_{1}t+k_{2})^{-\frac{3m}{k_{1}}}}{3k_{3}^{3}}+\frac{m}{%
k_{1}t+k_{2}}  \label{36}
\end{equation}%
\begin{equation}
H_{2}=\frac{(k_{1}t+k_{2})^{-\frac{3m}{k_{1}}-1}\left(
3k_{3}^{3}m(k_{1}t+k_{2})^{\frac{3m}{k_{1}}}-2c_{1}(k_{1}t+k_{2})\right) }{%
3k_{3}^{3}}  \label{37}
\end{equation}%
The expansion scalar $\theta $ and the shear $\sigma ^{2}$ are obtained as
\begin{equation}
\theta =3H=\frac{3m}{k_{1}t+k_{2}},\ \ \ \sigma ^{2}=\frac{%
c_{1}^{2}(k_{1}t+k_{2})^{-\frac{6m}{k_{1}}}}{3k_{3}^{6}}  \label{38}
\end{equation}%
The anisotropy parameter $\Delta $ of the expansion is
\begin{equation}
\Delta =6\left( \frac{\sigma }{\theta }\right) ^{2}=\frac{%
2c_{1}^{4}(k_{1}t+k_{2})^{2-\frac{12m}{k_{1}}}}{27k_{3}^{12}m^{2}}
\end{equation}%
\label{39} The other dynamical parameters for our model are obtained as
\begin{equation}
p=\frac{(k_{1}t+k_{2})^{-\frac{6m}{k_{1}}-2}\left(
\begin{array}{c}
2(3\alpha +2)c_{1}^{2}(k_{1}t+k_{2})^{2}-3c_{1}k_{3}^{3}m(k_{1}t+k_{2})^{%
\frac{3m}{k_{1}}+1} \\
+9k_{3}^{6}m(-2(2\alpha +1)k_{1}+6\alpha m+m)(k_{1}t+k_{2})^{\frac{6m}{k_{1}}%
}%
\end{array}%
\right) }{18\alpha (\alpha +1)k_{3}^{6}}  \label{40}
\end{equation}%
\begin{equation}
\rho =\frac{(k_{1}t+k_{2})^{-\frac{6m}{k_{1}}-2}\left(
\begin{array}{c}
2(\alpha +2)c_{1}^{2}(k_{1}t+k_{2})^{2}-3(2\alpha
+1)c_{1}k_{3}^{3}m(k_{1}t+k_{2})^{\frac{3m}{k_{1}}+1} \\
-9k_{3}^{6}m(2k_{1}+(4\alpha -1)m)(k_{1}t+k_{2})^{\frac{6m}{k_{1}}}%
\end{array}%
\right) }{18\alpha (\alpha +1)k_{3}^{6}}  \label{41}
\end{equation}%
\begin{equation}
\Lambda =\frac{(k_{1}t+k_{2})^{-\frac{6m}{k_{1}}-2}\left(
\begin{array}{c}
-2c_{1}^{2}(k_{1}t+k_{2})^{2}-3c_{1}k_{3}^{3}m(k_{1}t+k_{2})^{\frac{3m}{k_{1}
}+1} \\+9k_{3}^{6}m(2k_{1}-5m)(k_{1}t+k_{2})^{\frac{6m}{k_{1}}} \end{array} \right) }{
18(\alpha +1)k_{3}^{6}}  \label{42}
\end{equation}%
\begin{equation}
\omega =\frac{%
\begin{array}{c}
2(3\alpha +2)c_{1}^{2}(k_{1}t+k_{2})^{2}-3c_{1}k_{3}^{3}m(k_{1}t+k_{2})^{%
\frac{3m}{k_{1}}+1} \\ 
+9k_{3}^{6}m(-2(2\alpha +1)k_{1}+6\alpha m+m)(k_{1}t+k_{2})^{\frac{6m}{k_{1}}%
}%
\end{array}%
}{%
\begin{array}{c}
2(\alpha +2)c_{1}^{2}(k_{1}t+k_{2})^{2}-3(2\alpha
+1)c_{1}k_{3}^{3}m(k_{1}t+k_{2})^{\frac{3m}{k_{1}}+1} \\ 
-9k_{3}^{6}m(2k_{1}+(4\alpha -1)m)(k_{1}t+k_{2})^{\frac{6m}{k_{1}}}%
\end{array}%
}   \label{43}
\end{equation}

Finally, the metric (\ref{6}) reduces to

\begin{eqnarray}
ds^{2} &=&dt^{2}-c_{2}^{\frac{2}{3}}k_{3}^{2}(k_{1}t+k_{2})^{2\frac{m}{k_{1}}%
}\times \exp 2\left[ \frac{c_{1}(k_{1}t+k_{2})^{1-\frac{3m}{k_{1}}}}{%
3k_{3}^{3}(k_{1}-3m)}\right] (dx^{2}+dy^{2})  \notag \\
&&-c_{2}^{\frac{-4}{3}}k_{3}^{2}(k_{1}t+k_{2})^{2\frac{m}{k_{1}}}\times \exp
2\left[ \frac{-2c_{1}(k_{1}t+k_{2})^{1-\frac{3m}{k_{1}}}}{%
3k_{3}^{3}(k_{1}-3m)}\right] dz^{2}\text{,}  \label{44}
\end{eqnarray}

To have a better understanding of our obtained model, in the next section,
we take an example by constraining the model parameters with recent
observation and plot the cosmological parameters against cosmic time $t$.

\section{Exemplification}

From equation (\ref{33}) it is clear that for an accelerated expansion of
the Universe, we must have $k_{1}<m$. Recent observations suggested that the
numerical value of the deceleration parameter should lie in the range, $-%
\frac{1}{3}\leqslant q<0$ which will valid in our case if $\frac{2}{3}%
\leqslant \frac{k_{1}}{m}<0$. For a flat space-time, the parameters $k_{1}$,
$k_{2}$ and $m$ must satisfy the inequations $1.5\leq k_{1}\leq 3$, $2.5\leq
m\leq 4$ and $0<k_{2}<2$ \cite{PacifMishra15}. For an accelerated expansion
consistent with the observation, the numerical value of the deceleration
parameter at present may be $q_{p}\approx -0.55$. So, constraining the
values of $k_{1}$, $k_{2}$ and $m$ accordingly, we can study the evolution
of various cosmological parameters obtained in the previous section for our
obtained model. Looking at the range of these parameters, we choose here $%
k_{1}=1.59$, $m=3.59$, $k_{2}=0.7$ and see the evolution of these
cosmological parameters graphically as follows.

\begin{figure}[h]
\centering
\includegraphics[width=60mm]{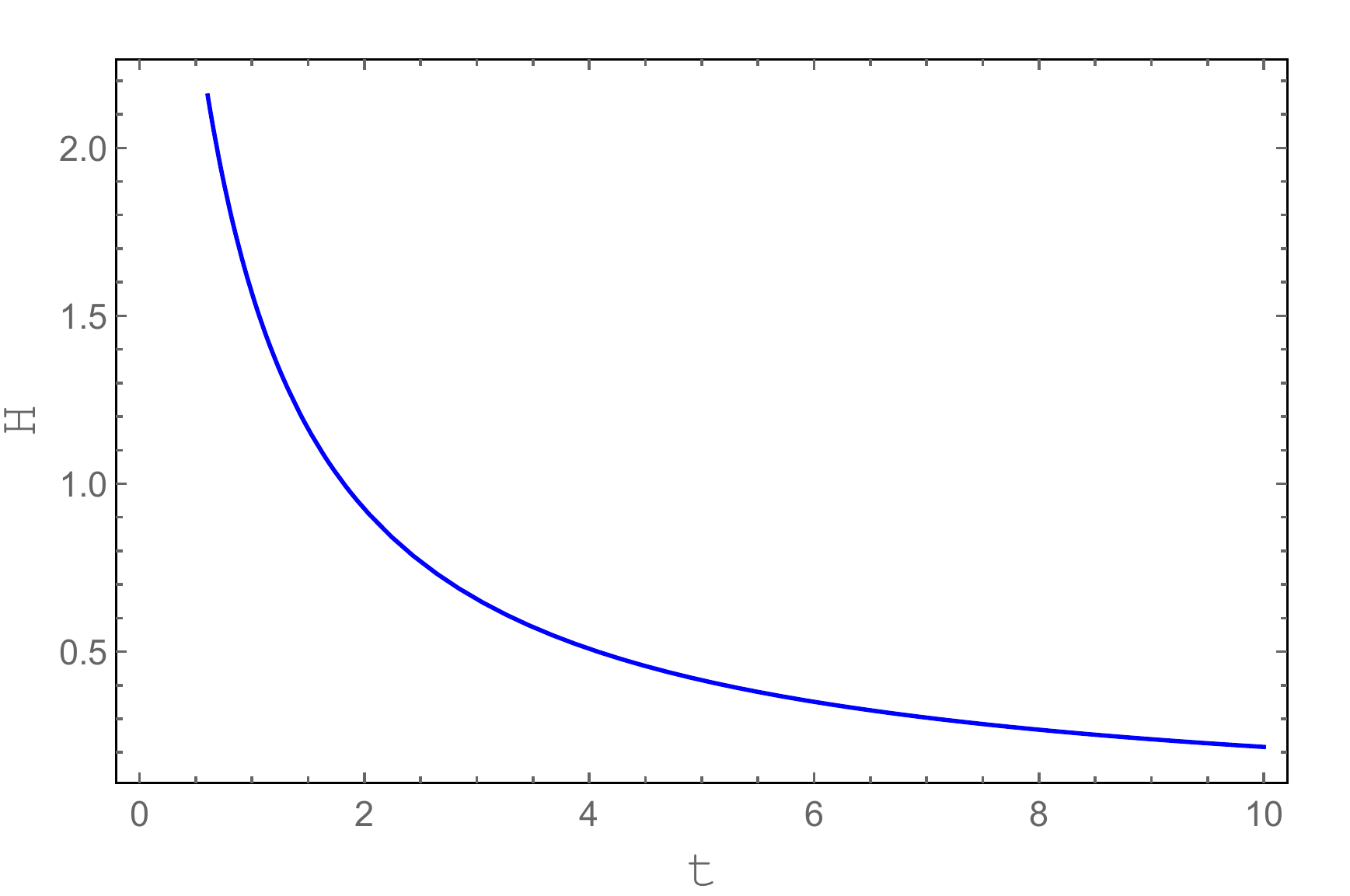}
\caption{Profile of Hubble parameter $(H)$ against time (in billion years) for $k_{1}=1.59$, $k_{2}=0.7$, $k_{3}=1$, $%
m=3.59$, $c_{1}=c_{2}=1$.}
\label{fig1}
\end{figure}

\begin{figure}[h]
\centering
\includegraphics[width=60mm]{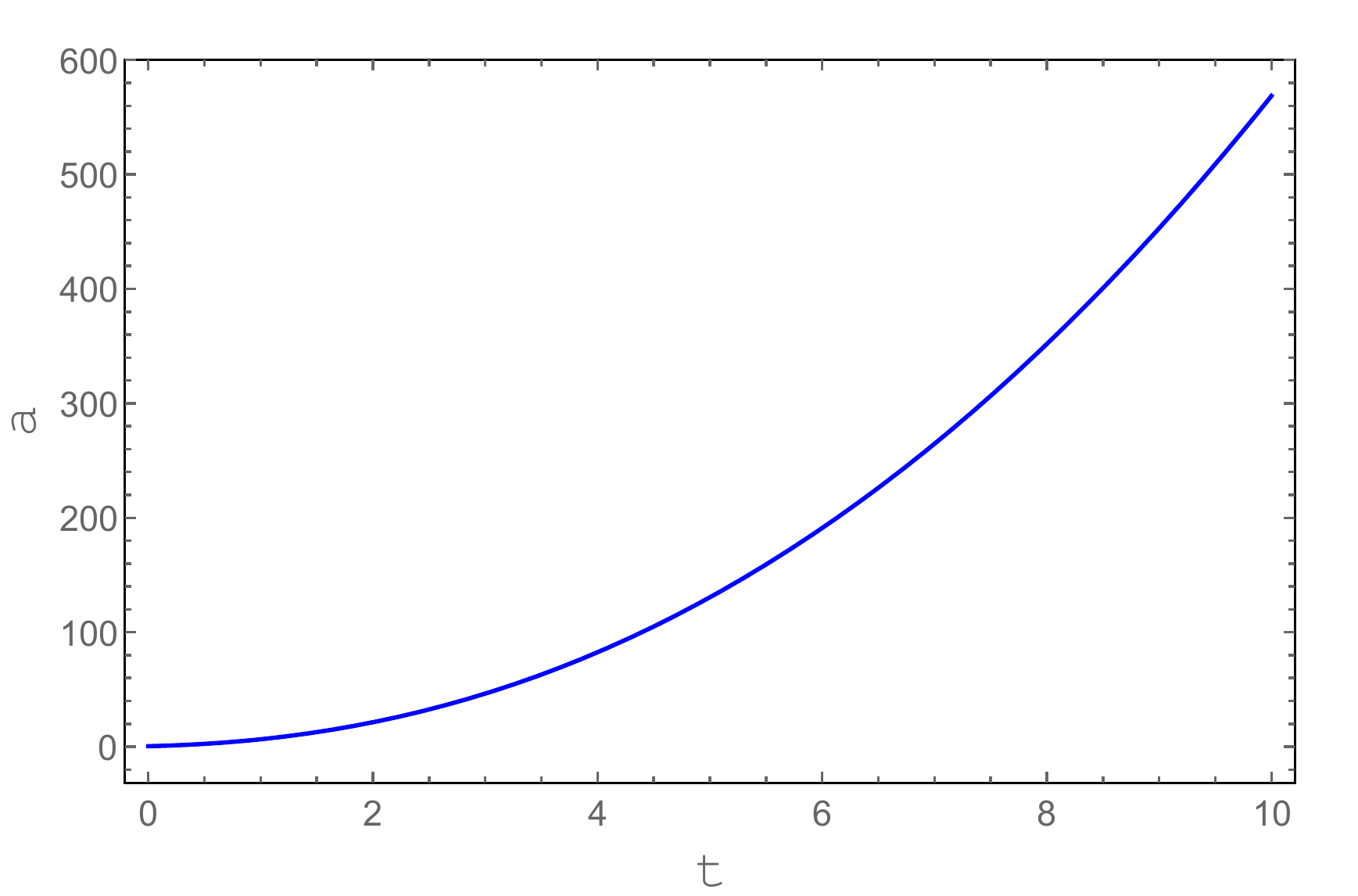}
\caption{Profile of Scale factor $(a)$ against time (in billion years) for $k_{1}=1.59$, $k_{2}=0.7$, $k_{3}=1$, $%
m=3.59$, $c_{1}=c_{2}=1$.}
\label{fig2}
\end{figure}

\begin{figure}[h]
\centering
\includegraphics[width=60mm]{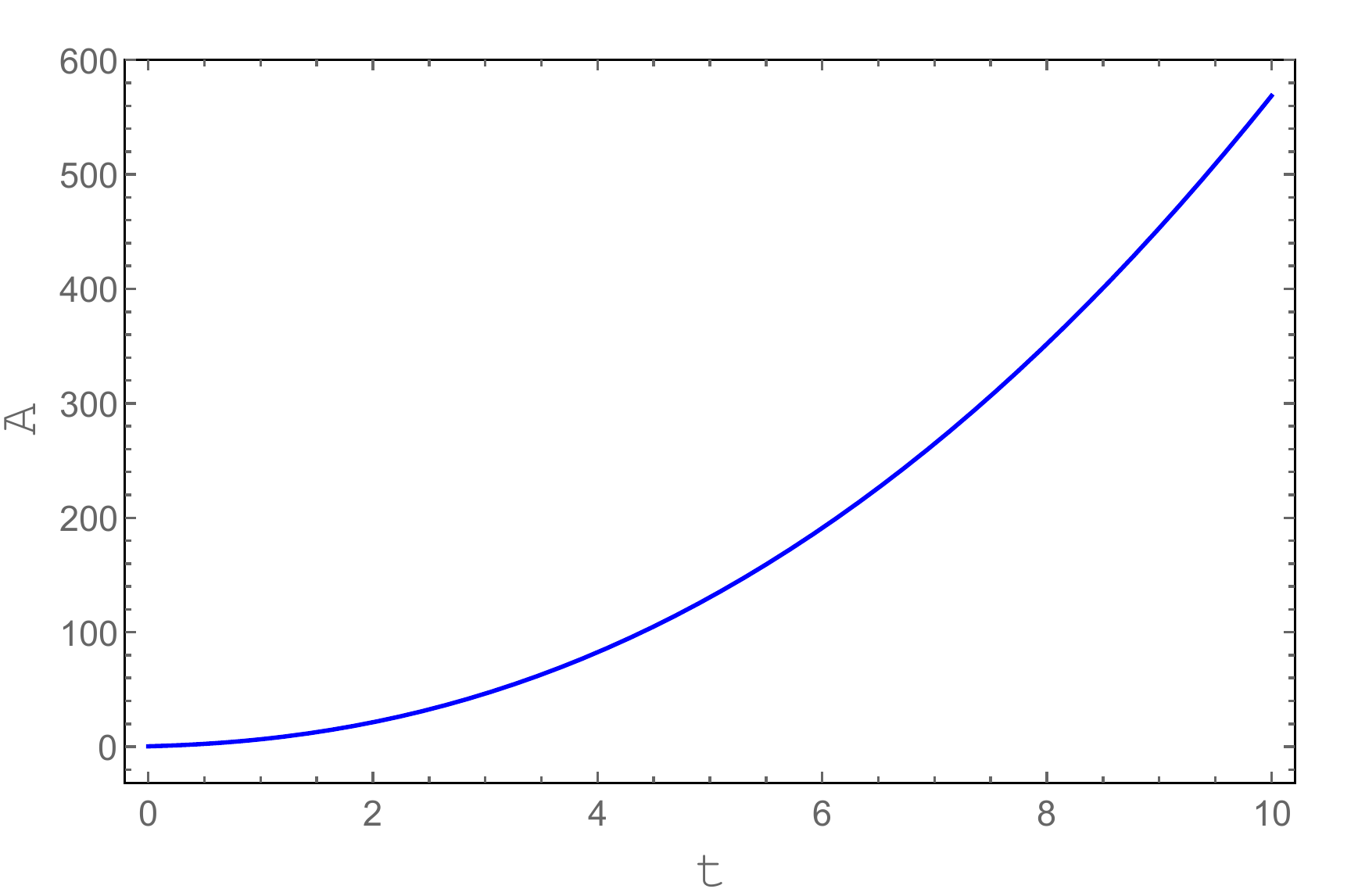}
\caption{Profile of Metric
potentials $A$ against time (in billion years) for $k_{1}=1.59$, $k_{2}=0.7$, $k_{3}=1$, $%
m=3.59$, $c_{1}=c_{2}=1$.}
\label{fig3}
\end{figure}

\begin{figure}[h]
\centering
\includegraphics[width=60mm]{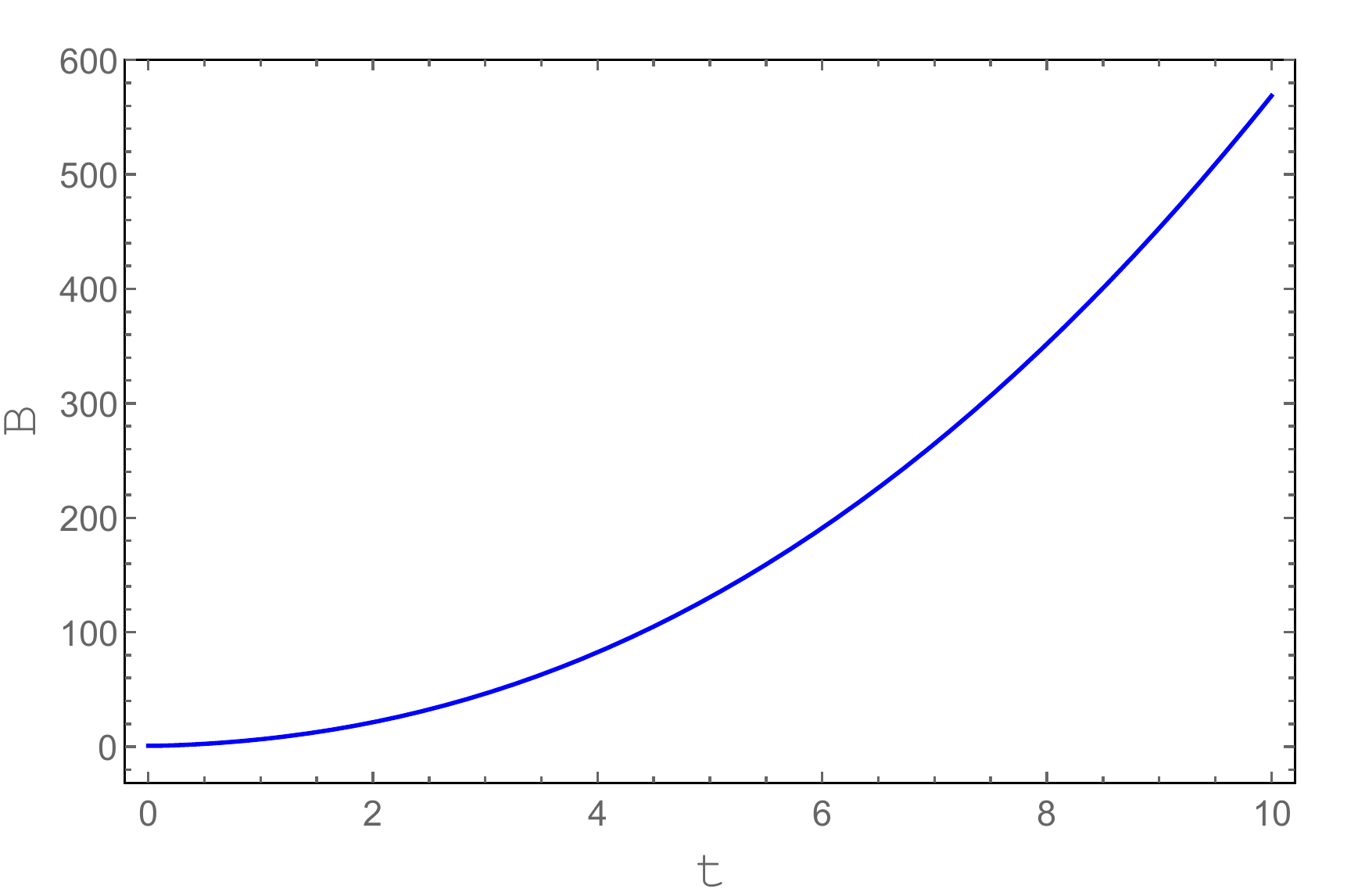}
\caption{Profile of Metric
potentials $B$ against time (in billion years) for $k_{1}=1.59$, $k_{2}=0.7$, $k_{3}=1$, $%
m=3.59$, $c_{1}=c_{2}=1$.}
\label{fig4}
\end{figure}

\begin{figure}[th]
\caption{Profile for energy density against time for $k_{1}=1.59$, $k_{3}=1$%
, $m=3.59$, $c_{1}=c_{2}=1$ for various values of $k_{2}$. }
\label{fig5}\centering
\minipage{0.50\textwidth} \includegraphics[width=60mm]{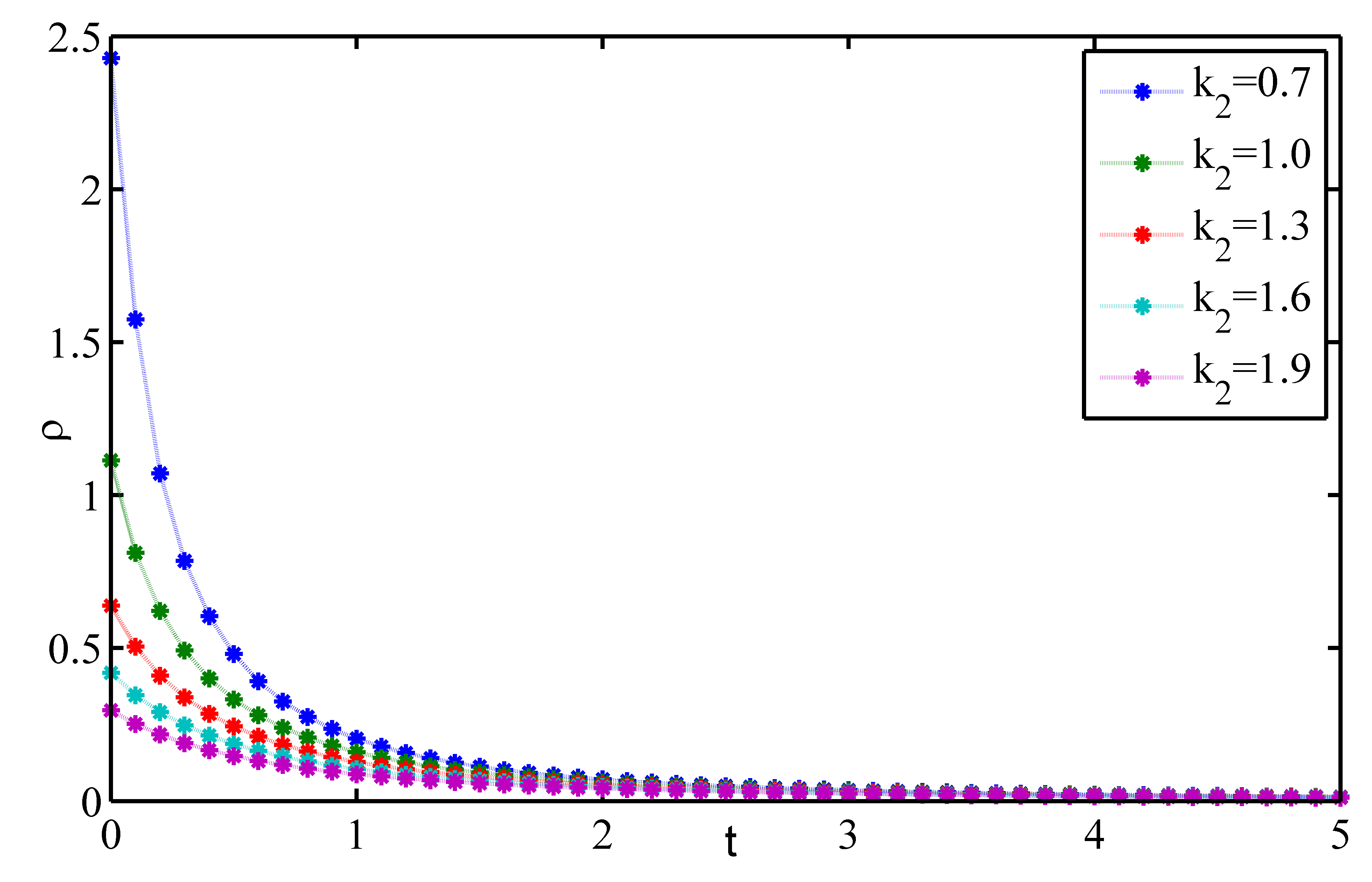} \endminipage%
\hspace{0.8cm} \minipage{0.50\textwidth} \includegraphics[width=60mm]{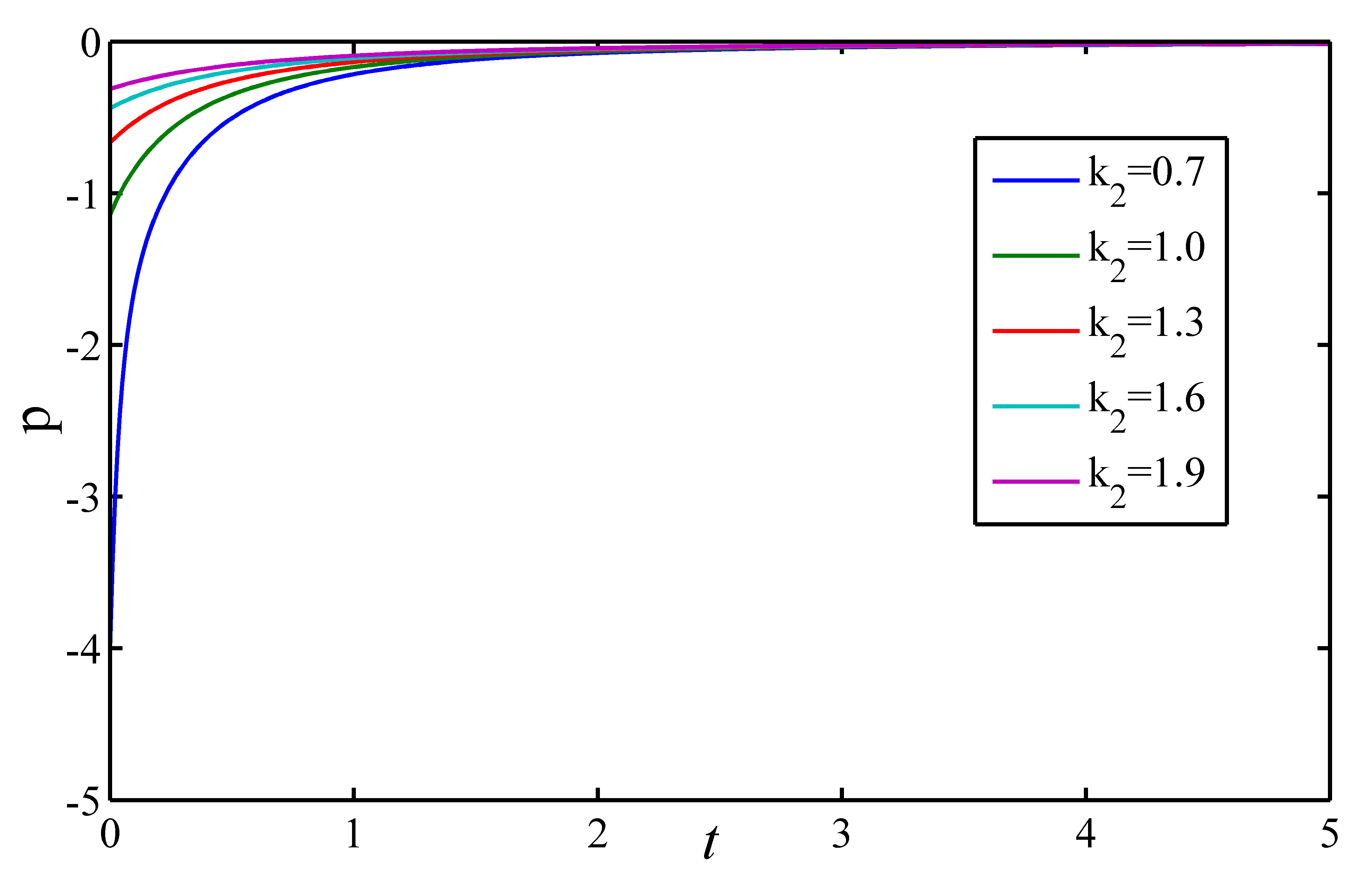}
\endminipage
\caption{Profile for pressure against time for $k_{1}=1.59$, $k_{3}=1$, $%
m=3.59$, $c_{1}=c_{2}=1$ for various values of $k_{2}$.}\label{fig6}
\end{figure}

\begin{figure}[ht]
\caption{Profile for cosmological constant against time for $k_1=1.59$, $%
k_3=1$, $m=3.59$, $c_1=c_2=1$ for various values of $k_2$.}
\label{fig7}\centering
\minipage{0.50\textwidth} \includegraphics[width=60mm]{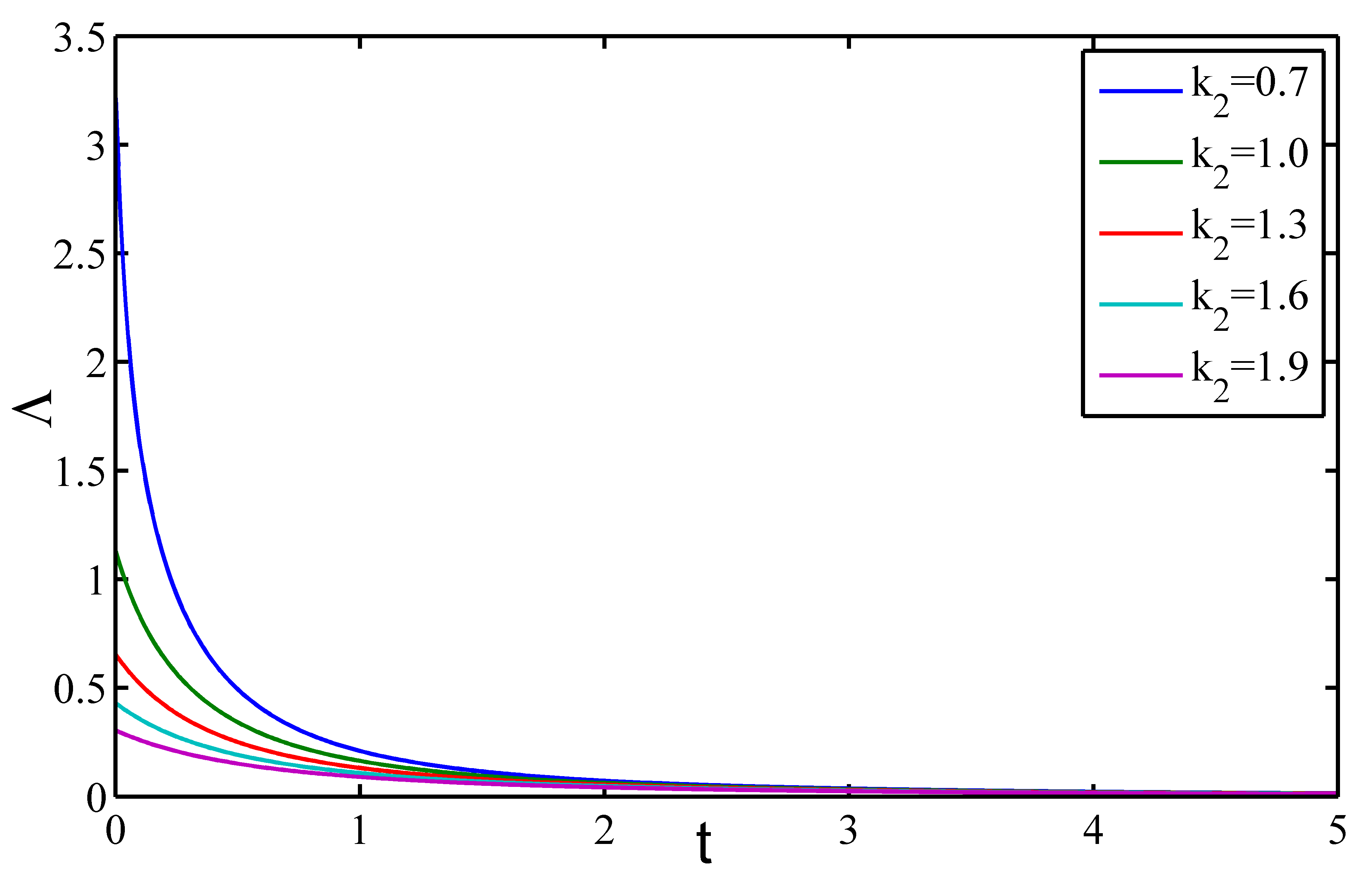} \endminipage%
\hspace{0.8cm} \minipage{0.50\textwidth} \includegraphics[width=60mm]{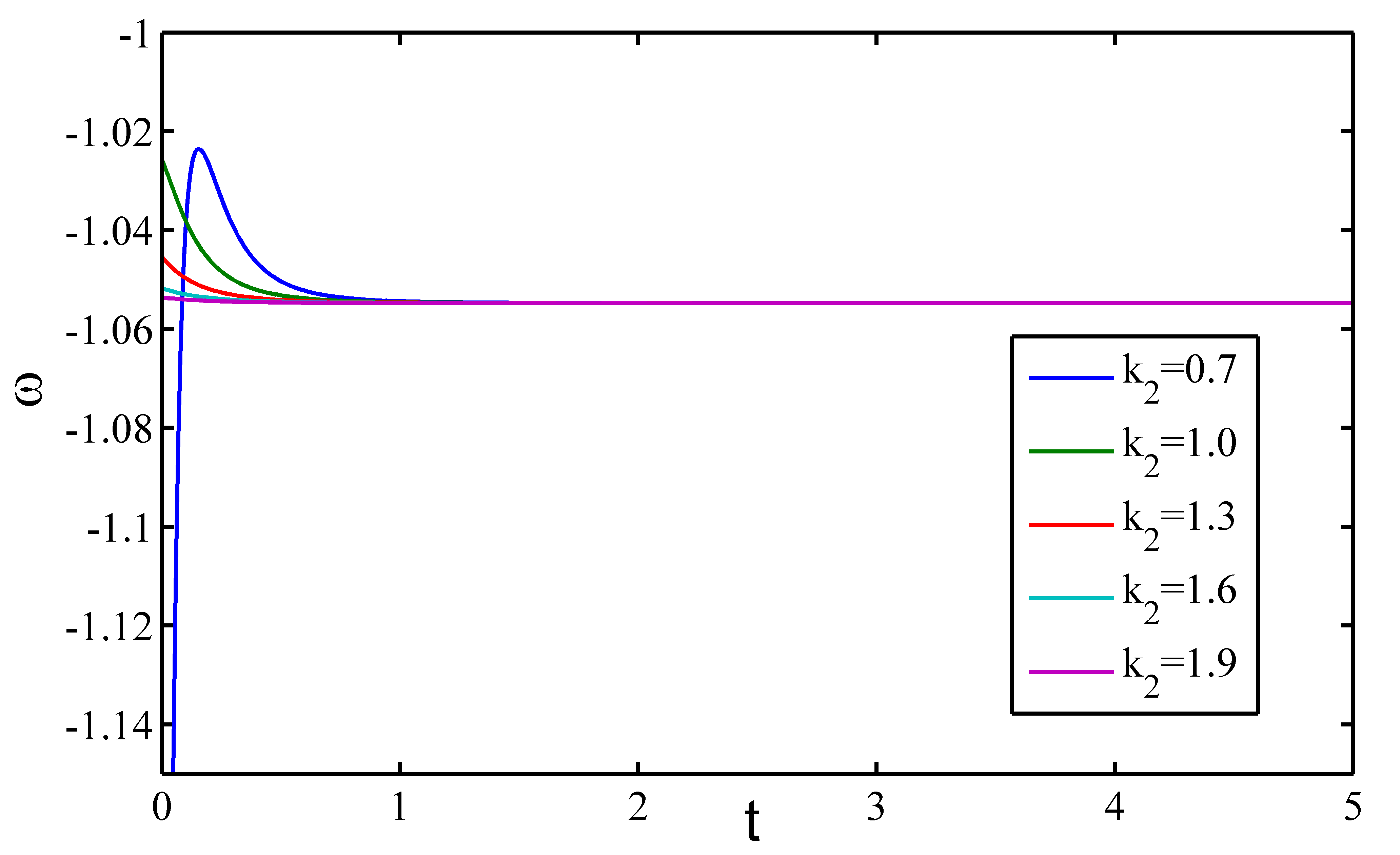}
\endminipage
\caption{Profile for EoS parameter against time for $k_1=1.59$, $k_3=1$, $%
m=3.59$, $c_1=c_2=1$ for various values of $k_2$.}
\label{fig8}
\end{figure}

The profile of Hubble parameter, scale factor and metric potentials are
presented in the Figures 1-4. Here we noticed from the Figure 1 that,
Hubble parameter is a decreasing function of time and it approaches towards
zero with the evolution of time. Scale factor and metric potentials are
increasing function of time and they are approaching to infinity with the
evolution of time i.e. $a,A,B\rightarrow \infty $ when $t\rightarrow \infty $%
. \newline
The profile of energy density and pressure is presented in the Figure \ref%
{fig5} and Figure \ref{fig6} respectively. Here we noticed from the figure
that, energy density is a decreasing function of time and it approaches
towards zero with the evolution of time. Here the positivity of energy
density, tighten the interval of $k_{2}$ from $0<k_{2}<2$ to $0.6<k_{2}<2$.
The pressure of the model is also approaching to zero with the evolution of
time and it is negative, which follow the observational data. \newline
Figure \ref{fig7} and Figure \ref{fig8} represents the profile of
cosmological constant and EoS parameter against time. The cosmological
constant is positive and decreasing function of time. Here $\Lambda
\rightarrow 0$ when $t\rightarrow \infty $. The EoS parameter is negative
valued function and which is less than $-1$. It means that, our models
represents the phantom energy cosmological model.

\begin{figure}[ht]
\caption{Profile for Look-back time against red-shift for $m=3.59$, $H_0=60$
for various values of $k_1$.}
\label{fig9}\centering
\minipage{0.50\textwidth} \includegraphics[width=60mm]{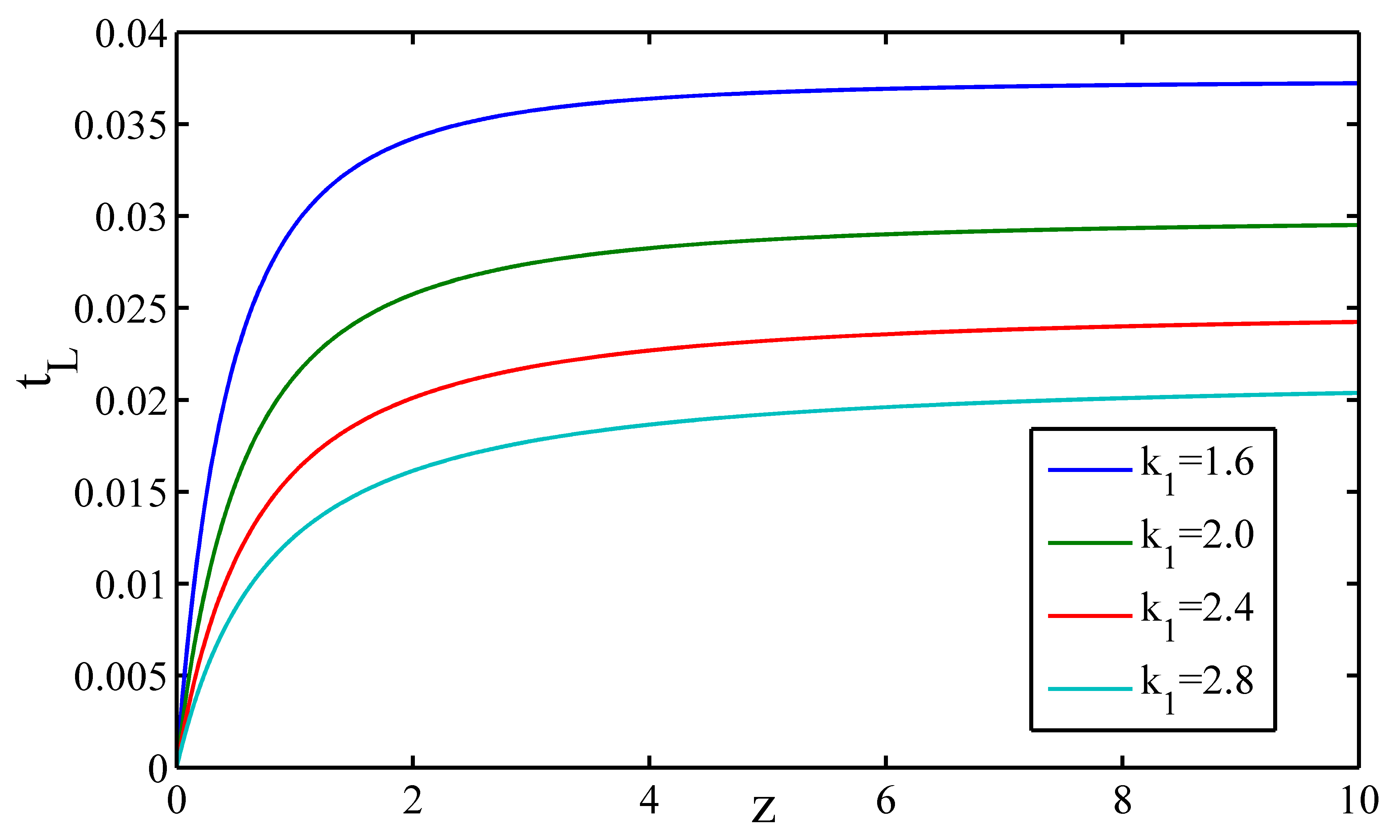} %
\endminipage\hspace{0.8cm} \minipage{0.50\textwidth} %
\includegraphics[width=60mm]{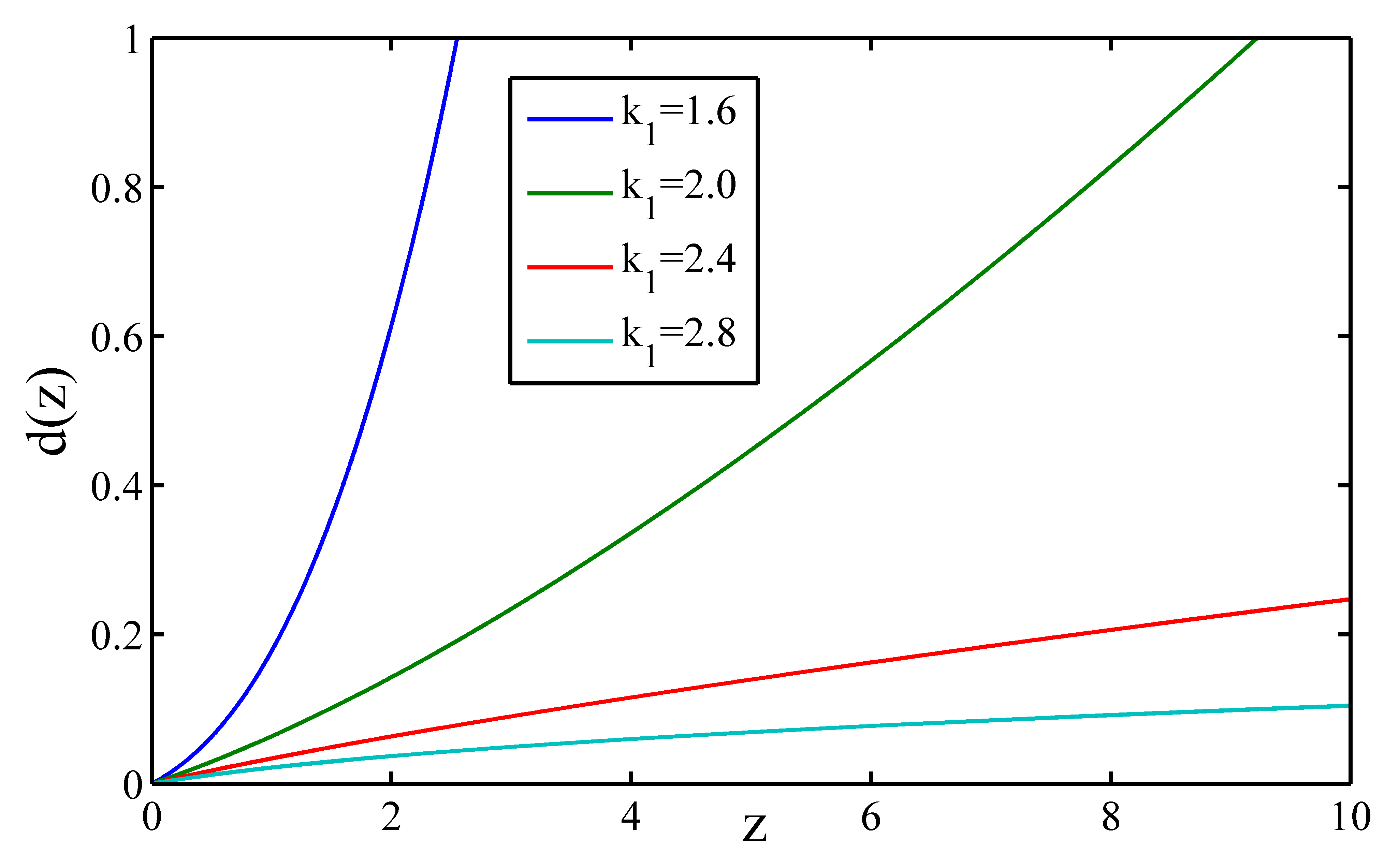} \endminipage
\caption{Profile for Proper distance against red-shift for $m=3.59$, $H_0=60$%
, $k_3=1$ for various values of $k_1$.}
\label{fig10}
\end{figure}

\begin{figure}[ht]
\caption{Profile for Luminosity distance against red-shift for $m=3.59$, $%
H_0=60$, $k_3=1$ for various values of $k_1$.}
\label{fig11}\centering
\minipage{0.50\textwidth} \includegraphics[width=60mm]{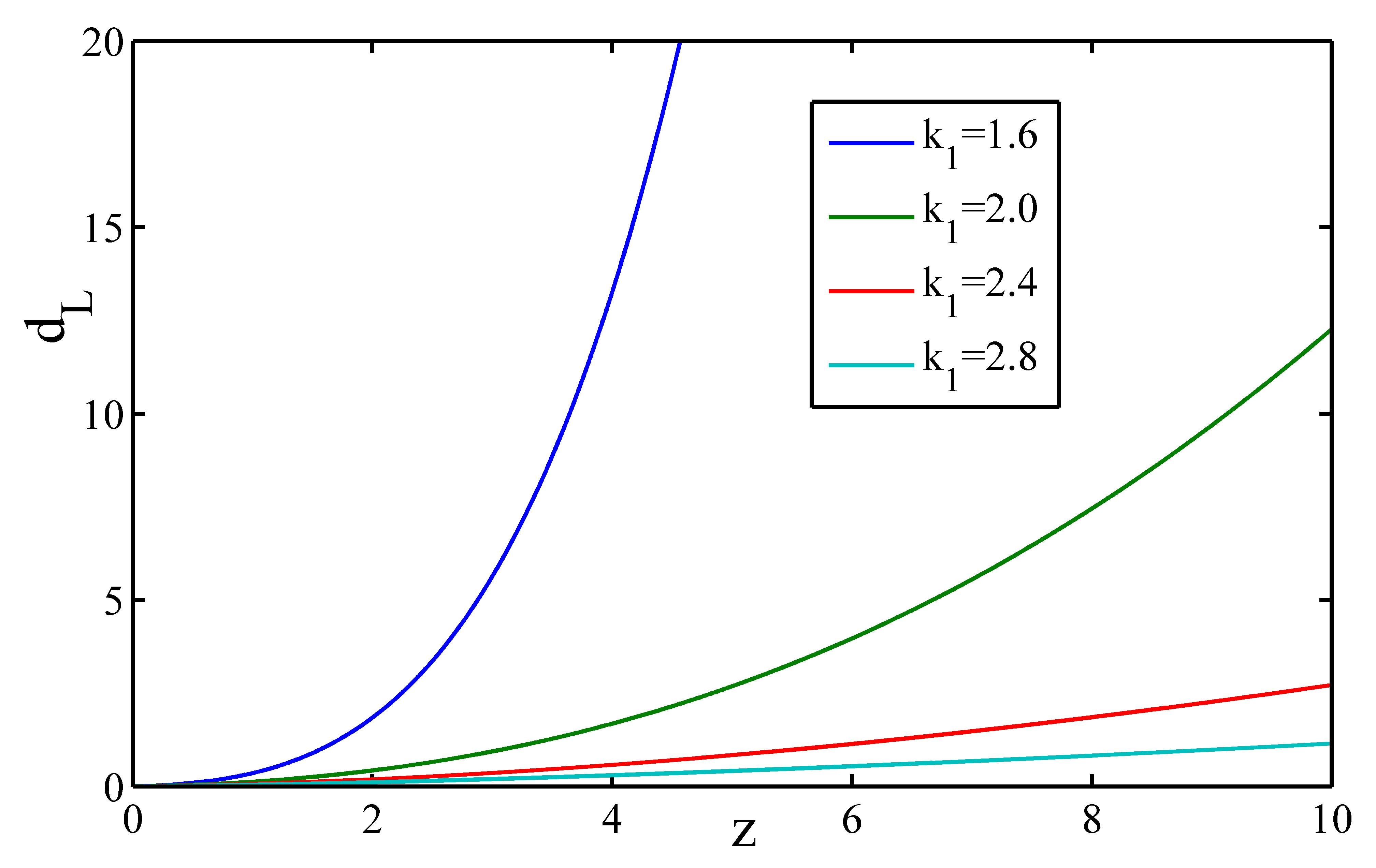} %
\endminipage\hspace{0.8cm} \minipage{0.50\textwidth} %
\includegraphics[width=60mm]{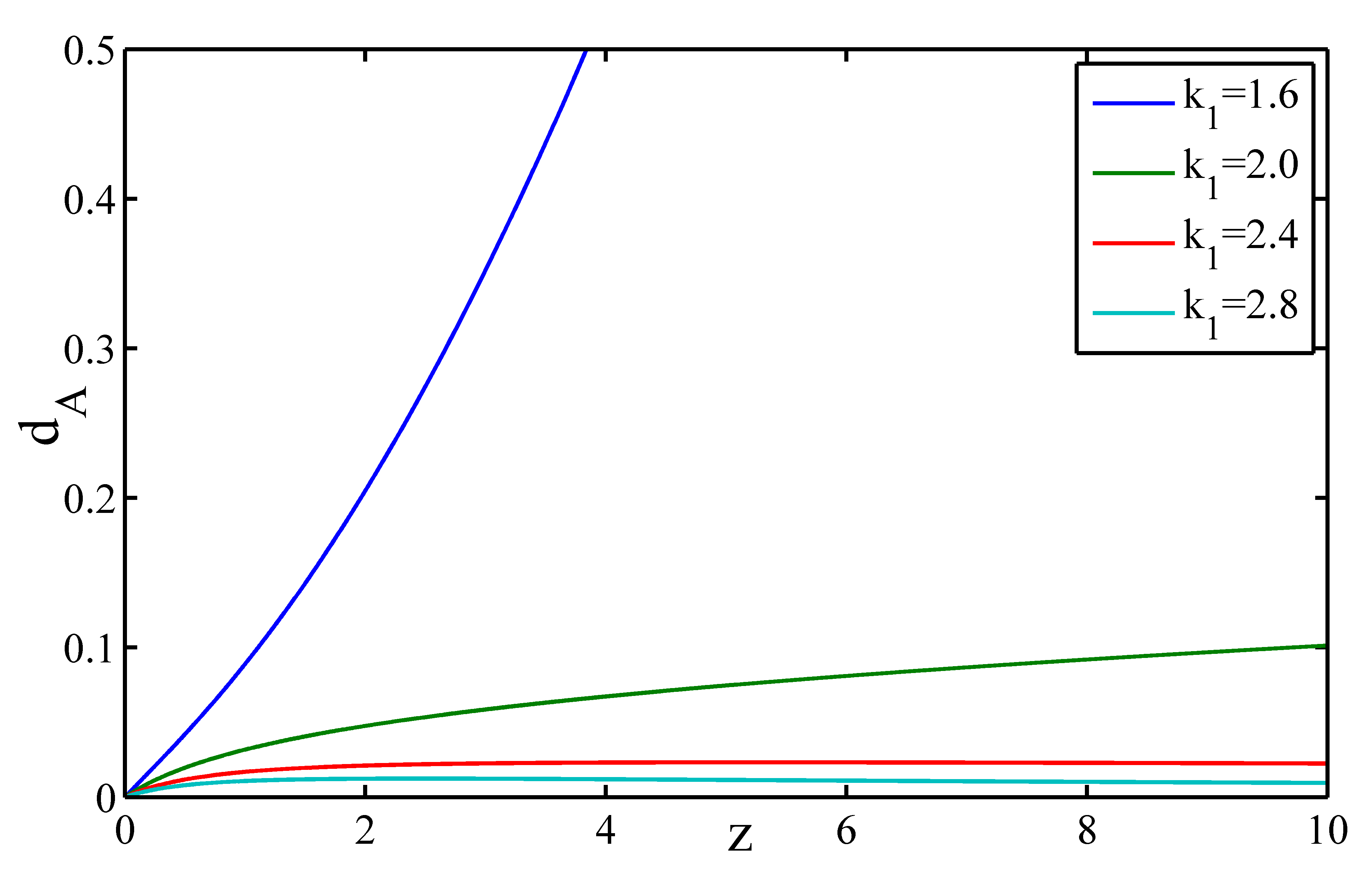} \endminipage
\caption{Profile for Angular-diameter distance against red-shift for $m=3.59$%
, $H_0=60$, $k_3=1$ for various values of $k_1$.}
\label{fig12}
\end{figure}

\section{Distances in Cosmology}

Distance is one of the basic measurement that we can performed. In the
history of astronomy, distance measurement played a important role and some
time surprising role for understanding about Universe. In this section, we
have presented some of the different distance measures.

\subsection{Look-back time-redshift}

The look-back time $t_{L}$ is defined as the difference between the present
age of the Universe $t_{0}$ and the age of the Universe, when a particular
light from a cosmic source at a particular redshift $z$ was emitted. Thus it
is defined as
\begin{equation}
t_{L}=t_{0}-t(z)=\int_{a}^{a_{0}}\frac{da}{\dot{a}},  \label{e45}
\end{equation}%
where $a_{0}$ is the present day scale factor of the Universe. The scale
factor of the Universe $a(t)$ is related to $a_{0}$ by the relation
\begin{equation}
\frac{a}{a_{0}}=\frac{1}{1+z}  \label{e46}
\end{equation}%
For the discussed model, we have
\begin{equation}
k_{1}t+k_{2}=(k_{1}t_{0}+k_{2})(1+z)^{-\frac{m}{k_{1}}}  \label{e47}
\end{equation}%
The above equation takes the form
\begin{equation}
H_{0}(t_{0}-t)=\frac{m}{k_{1}}\left[ 1-(1+z)^{-\frac{m}{k_{1}}}\right]
\label{e48}
\end{equation}%
Here $H_{0}$ is the Hubble constant at present. The value of $H_{0}$ is lies
between $50-100$ km s$^{-1}$ Mpc$^{-1}$. The equation (48) can also be
expressed as
\begin{equation}
H_{0}(t_{0}-t)=\left( \frac{m}{k_{1}}\right) ^{2}\left[ z-\frac{m+k_{1}}{%
2k_{1}}z^{2}+\frac{(m+k)(m+2k)}{6k^{2}}z^{3}-\cdot \cdot \cdot \cdot \cdot %
\right]  \label{e49}
\end{equation}%
With the help of $q=-1+\frac{k_{1}}{m}$, equation (49) takes the form
\begin{equation}
H_{0}(t_{0}-t)=\frac{1}{(1+q)^{2}}\left[ z-\frac{2+q}{2(1+q)}z^{2}+\frac{%
(2+q)(3+2q)}{6(1+q)}z^{3}-\cdot \cdot \cdot \cdot \cdot \right]  \label{e50}
\end{equation}%
When $z\rightarrow \infty $, equation (48) reads
\begin{equation}
t_{L}=t_{0}-t=\frac{m}{k_{1}}H_{0}^{-1}=\frac{H_{0}^{-1}}{1+q}  \label{e51}
\end{equation}%
For small $z$, $H_{0}(t_{0}-t)$ can be approximated as
\begin{equation}
H_{0}(t_{0}-t)\approx \left( \frac{m}{k_{1}}\right) ^{2}z=\frac{z}{(1+q)^{2}}
\label{e52}
\end{equation}

\subsection{Proper Distance}

The proper distance $d(z)$ is defined as the distance between a cosmic
source emitting light at any instant $t=t_1$, located at $r=r_1$ with
redshift $z$ and the observer receiving the light from the source emitted at
$r=0$ and $t=t_0.$ Thus
\begin{equation}  \label{e53}
d(z)=r_1a_0,\; \text{where} r_1=\int_{t_1}^{t_0}\frac{dt}{a}
\end{equation}
For the discussed model, we have the proper distance as
\begin{equation}  \label{e54}
d(z)=\frac{mH_0^{-1}}{k_3(k_1-m)}\left[1-(1+z)^{-\frac{m}{k_1}(1-\frac{m}{k_1%
})}\right]
\end{equation}
The above expression indicates that, $d(z)\rightarrow \frac{mH_0^{-1}}{%
k_3(k_1-m)}$ when $z\rightarrow \infty$ for $0<\frac{m}{k_1}<1$ and $%
d(z)\rightarrow \infty$ when $z\rightarrow \infty$ for $\frac{m}{k_1}\geq 1$.

\subsection{Luminosity distance}

The apparent luminosity of a source at radial coordinate $r_{1}$ with a
redshift $z$ of any size $l$ is defined as
\begin{equation}
l=\frac{L}{4\pi r_{1}^{2}a_{0}^{2}(1+z)^{2}},  \label{e55}
\end{equation}%
where $L$ is the absolute luminosity distance. Let us introduce a luminosity
distance $d_{L}$ as
\begin{equation}
d_{L}=\left( \frac{L}{4\pi l}\right) =a_{0}r_{1}(1+z)  \label{e56}
\end{equation}%
With the help of equation (53), equation (56) takes the form
\begin{equation}
d_{L}=d(z)(1+z)  \label{e57}
\end{equation}%
For the discussed model, we have Luminosity distance $d_{L}$ as
\begin{equation}
d_{L}=\frac{mH_{0}^{-1}}{k_{3}(k_{1}-m)}\left[ 1-(1+z)^{-\frac{m}{k_{1}}(1-%
\frac{m}{k_{1}})}\right] (1+z)  \label{e58}
\end{equation}

\subsection{Angular-diameter distance}

The angular-diameter distance $d_A$ is defined such that
\begin{equation*}
\theta=\frac{l}{d_A},
\end{equation*}
where $\theta$ is the angle subtended by an object of size $l$. It is also
defined in term of proper distance and Luminosity distance as
\begin{equation*}
d_A=d(z)(1+z)^{-1}=d_L(1+z)^{-2}.
\end{equation*}
For the presented model
\begin{equation}  \label{e59}
H_0d_A=\frac{m}{k_3(k_1-m)}\left[1-(1+z)^{-\frac{m}{k_1}(1-\frac{m}{k_1})}%
\right](1+z)^{-1}.
\end{equation}

\subsection{Distance Modulus}

The distance modulus $(\mu(z))$ is given as
\begin{equation*}
\mu(z)=5\log d_L+25
\end{equation*}%
Thus, the distance modulus $(\mu(z))$ in terms of redshift parameter $z$
is obtained as
\begin{equation}  \label{e60}
\mu(z)=5\log\left(\frac{mH_0^{-1}}{k_3(k_1-m)}\left[1-(1+z)^{-\frac{m}{k_1}%
(1-\frac{m}{k_1})}\right]\right)+25
\end{equation}

\section{Conclusion}

In this article, we have presented a new solution to the field equations by
using the law of variation for Hubble's parameter which yield constant
deceleration parameter. The law of variation for Hubble parameter in Eq.
(31) explicitly determine the values of the scale factors. One can solve
Einstein field equations for Bianchi type metric with this functional form
of Hubble parameter in principle. For $k_{1}=0$, the deceleration parameter $%
q=-1$ and $\frac{dH}{dt}=0$, which gives the greatest value of $H$ and
fastest rate of expansion as presented in Figure 1. This type of solutions
are consistent as per the recent observations for an accelerated expansion
of the universe. The variation of Hubble parameter presented in this paper
may be used to study new solutions of Einstein field equations in modified
theories of gravity. The model obtained in Eq. (44) of the universe start
with a singularity at $t=-\frac{k_{2}}{k_{1}}$ and remain regular in finite
region. The expansion rate goes down with time and finally tend to zero as $%
t\rightarrow \infty $. From the anisotropy parameter, it is observed that
the model of the universe remains anisotropic throughout the evolution. The
energy density approaches to zero as $t\rightarrow \infty $. The EoS
parameter clearly shows that this model is in phantom region. Finally, we
have discussed the consistency of this model with the distance parameters
such as look back time, proper distance, luminosity distance, angular
diameter distance and the distance modulus (see Figure \ref{fig9} to Figure %
\ref{fig12}).

\section{Acknowledgements}
Author SKJP wish to thank National Board of
Higher Mathematics, Department of Atomic Energy (DAE), Government of India
for financial support through post doctoral research fellowship. Author PKS
wish to thank M. Sami for his support and CTP, JMI for hospitality where a
part of this work have been done. We are very indebted to the  editor and the anonymous referee for illuminating suggestions that have significantly
improved our paper in terms of  research  quality as well as presentation.

\end{document}